\documentclass[12pt]{article}
\usepackage{graphicx}
\usepackage{graphics}
\usepackage[margin=.8in]{geometry}
\usepackage{amsmath,amssymb}
\usepackage{hyperref}
\usepackage{siunitx}
\usepackage{upgreek}	
\usepackage{authblk}
\usepackage[font=small,labelfont=bf]{subcaption}
\usepackage[font=small,labelfont=bf]{caption}

\begin{document}
		
		\title{On the presence of a critical detachment angle in gecko spatula peeling - A numerical investigation using an adhesive friction model}
		\author[1]{\small{Saipraneeth Gouravaraju}}
		\affil[1]{\footnotesize{Indian Institute of Technology Guwahati, Guwahati, India 781039}}
		\author[2,3]{\small{Roger A. Sauer}}
		\affil[2]{\footnotesize{Graduate School, Aachen Institute for Advanced Study in Computational Engineering Science (AICES), RWTH Aachen University, Templergraben 55, 52056 Aachen, Germany}}
		\affil[3]{\footnotesize{Department of Mechanical Engineering, Indian Institute of Technology Kanpur, UP 208016, India}}
		\author[1]{\small{Sachin Singh Gautam}\footnote{Corresponding Author, email: \href{mailto:ssg@iitg.ac.in}{ssg@iitg.ac.in}}}
		\date{}
		\maketitle
		\begin{abstract}
			A continuum-based computational contact model is employed to study coupled adhesion and friction in gecko spatulae. Nonlinear finite element analysis is carried out to simulate spatula peeling from a rigid substrate. It is shown that the ``frictional adhesion" behavior, until now only observed from seta to toe levels, is also present at the spatula level. It is shown that for sufficiently small spatula pad thickness, the spatula detaches at a constant angle known as the critical detachment angle irrespective of the peeling and shaft angles. The spatula reaches the same energy states at the jump-off contact point, which directly relates to the invariance of the critical detachment angle. This study also reveals that there is an optimum pad thickness associated with the invariance of the critical detachment angle. It is further observed that the sliding of the spatula pad is essential for the invariance of the critical detachment angle.  
		\end{abstract}
	    \textbf{keywords}: Gecko adhesion, nonlinear finite element analysis, computational contact, adhesive friction, critical detachment angle.
	    \section{Introduction}
	    Geckos employ microscopic hairs (so-called setae) on their toe pads to reversibly adhere to any kind of substrate. Each foot has about five hundred thousand setae \cite{Autumn2000}. Each seta contains hundreds of spatula-like projections which make intimate contact with the substrate. The experiments of Autumn et al. \cite{Autumn2000, Autumn2002b} and Autumn and Peattie \cite{Autumn2002a} on an isolated Tokay Gecko seta have lead to important revelations in the understanding of the underlying mechanism of attachment and detachment. It was shown that a single seta of a Tokay Gecko can generate a frictional force of $200$ $\upmu$N and an adhesive normal force of $20-40$ $\upmu$N. Further, experiments by various researchers \cite{Autumn2002a,Huber2005a,Sun2005,Autumn2006a} revealed that, although capillary forces contribute towards the adhesion, the primary source of adhesion is widely accepted as short-range van der Waals forces.
	    
	    In the past 20 years, researchers have made great strides in understanding the gecko adhesive system using various experimental \cite{Russell2002,Autumn2007,Pugno2008a}, analytical \cite{Kwak2010,Peng2016,Labonte2019} as well as numerical techniques \cite{Peng2010,Sauer2009b,Sauer2010,Sauer2011,Gautam2014}. This knowledge has been applied to design different engineering applications \cite{Asbeck2009,Drotlef2017,Hou2018}  One of the remarkable and most researched properties of the adhesive structures of geckos is their ability to generate high attachment forces in their stance phase of the stride and swift detachment from the substrates during swing phase. It has been observed that despite generating surprisingly large forces, geckos can detach their toes in just $15-20$ ms and with negligible force \cite{Autumn2007}. In their experiments, Autumn et al. \cite{Autumn2000} observed that the seta was able to detach with negligible force just by changing the angle that the seta shaft makes with the substrate to $30^\circ$. The analytical model of Sitti and Fearing \cite{Sitti2003} and the finite element simulations of Gao et al. \cite{Gao2005} also revealed that increasing the shaft angle above $30^\circ$ causes breaking of the adhesive bonds between seta and the substrate. 
	    
	    Gecko setae and setal arrays are naturally curved proximally, i.e. towards the animal \cite{Autumn2002a,Russell2002}. Autumn et al. \cite{Autumn2006b} observed that proximal drag of the gecko setae and setal arrays resulted in tensile loading of the seta, and yet under these tensile normal loads setae could generate strong static and kinetic friction. This behavior is unlike any of the other materials and seems to violate Amontons' law of friction. Further, it was observed that at each level in the hierarchy of the gecko adhesive system (toes, arrays of seta, and seta), the adhesive structures detached from the substrate when the angle of the resultant force reached a particular value called the \emph{critical detachment angle}. This critical detachment angle varied among the hierarchical levels but is found to be constant for a particular level. Based on these observations, Autumn et al. \cite{Autumn2006b} proposed a phenomenological model called ``frictional adhesion" in which the normal adhesion is limited by the frictional force and the critical detachment angle. According to this model, geckos can stay attached to the substrate only if the angle between adhesive and frictional forces is less than the critical detachment angle.
	    
		A considerable amount of experimental, analytical as well as numerical research has been conducted to understand the relationship between adhesion and friction. Majidi et al. \cite{Majidi2006} used a generalized Amontons' law that accounts for tensile adhesive forces in order to study the adhesion and friction of microfiber arrays. Tian et al. \cite{Tian2006} estimated the frictional and adhesion forces developed by a single spatula according to the frictional adhesion model of Autumn et al. \cite{Autumn2006b}. Chen et al. \cite{Chen2009}, proposed a modified Kendall peeling model \cite{Kendall1975} and examined a hypothesis that parallel drag following the perpendicular preloading of setae causes pre-tension to build up in the spatulae. They found that, beyond a value of pre-tension, the pull-off force drops at a critical angle, where the detachment occurs independently of the pull-off force. Schubert et al. \cite{Schubert2008} studied the effect of sliding on the adhesion of microfiber arrays such as gecko setae and spatulae. The authors found that the attachment and detachment of the fibers are controlled by the shear force, which is consistent with the experimental observation in geckos \cite{Autumn2006b}. Jagota and Hui \cite{Jagota2011} reviewed the existing literature on gecko adhesion to understand the ways in which the adhesion, friction and the anisotropy of the gecko adhesive structures is modeled and analyzed. They also proposed an analytical model for sliding friction. Cheng et al. \cite{Cheng2012} performed numerical simulations to demonstrate that a non-uniform pre-tension develops in the spatulae during attachment and detachment.  Begley et al. \cite{Begley2013} studied single and double sided peeling of an elastic tape using a frictional sliding model similar to that of Jagota and Hui \cite{Jagota2011}. Their model shows that frictional sliding requires a higher critical force than during pure sticking. Labonte and Federle \cite{Labonte2016} examined the shear-sensitive adhesion in climbing animals. They observed that for high peeling angles $>30^\circ$, the pull-off forces were consistent with the classical analytical models such as Kendall's peeling model. But at peeling angles $<30^\circ$ the pull-off forces increased dramatically. To study friction due to adhesion, Mergel et al. \cite{Mergel2018} introduced two continuum-based contact models based on the coarse-grained contact model of Sauer and Li \cite{RogerLi2007}. The first model, named ``Model DI" defines a sliding threshold independent of the normal distance between the interacting surfaces, while the second model, named ``Model EA", defines a sliding threshold dependent on the normal distance. Using ``Model EA", Gouravaraju et al. \cite{Gouravaraju2020} recently investigated the spatula pull-off behavior and the influence of various geometric and material parameters on the pull-off forces. They have shown that as the spatulae are pulled at small angles, the shear force increases, which in turn increases the adhesion force. Mergel et al. \cite{Mergel2020} have given a general 3D computational framework of the continuum contact models in \cite{Mergel2018} and discussed its application to various adhesive contact problems. 
	    
	    Based on the literature, it can be observed that the interaction of friction and adhesion at the spatula level is still not studied in detail due to the difficulty in isolating a single spatula. Hence, the main objective of the present work is to investigate the possibility of the presence of adhesive friction\footnote[1]{the present authors prefer the term ``adhesive friction" instead of ``frictional adhesion" in order to describe the interaction between friction and adhesion as friction depends on adhesion (and not vice versa) according to the coupled adhesion-friction model used here (see Eq.~(\ref{eq:Sliding_Threshold})).} at the spatula level. Another objective is to explore the existence of a critical detachment angle at the spatula level, which has been shown to exist in the theoretical results of Chen et al.~\cite{Chen2009}. The influence of various parameters such as the peeling angle, the shaft angle, and the spatula pad thickness on the critical detachment angle is investigated. An explanation for the invariance of the critical detachment angle is sought by analyzing the energy and pull-off force evolution during the peeling process.
	    
	    \section{Adhesive friction formulation}
\begin{figure}[h!]
	    	\begin{center}
	    		\includegraphics[scale=0.5]{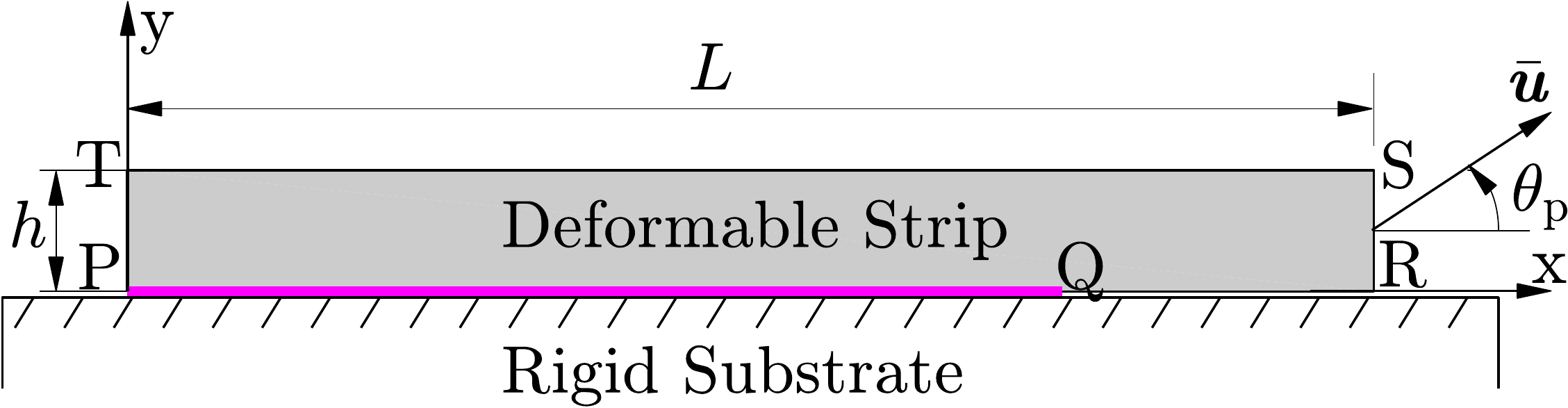}
	    		\caption{A deformable strip on a rigid substrate. The displacement $\bar{\boldsymbol{u}}$ is applied at angle of $\theta_\mathrm{p}$ called the peeling angle.}
	    		\label{fig:strip_orig}
	    	\end{center}
	    \end{figure}

	    A thin two-dimensional strip with dimensions $L \times h$ is used to model the gecko spatula, see Figure~\ref{fig:strip_orig} \cite{Tian2006,Peng2010,Chen2009}. It is assumed that only $75\%$ of the bottom surface (``PQ") interacts with the rigid substrate through intermolecular van der Waals forces \cite{Israelachvili_book}. As such, ``PQ" is considered as the spatula pad and ``QR" as the spatula shaft. 
	    
	    \begin{figure}[h!]
	    	\begin{center}
	    		\hspace{0.7cm} \includegraphics[scale=0.5]{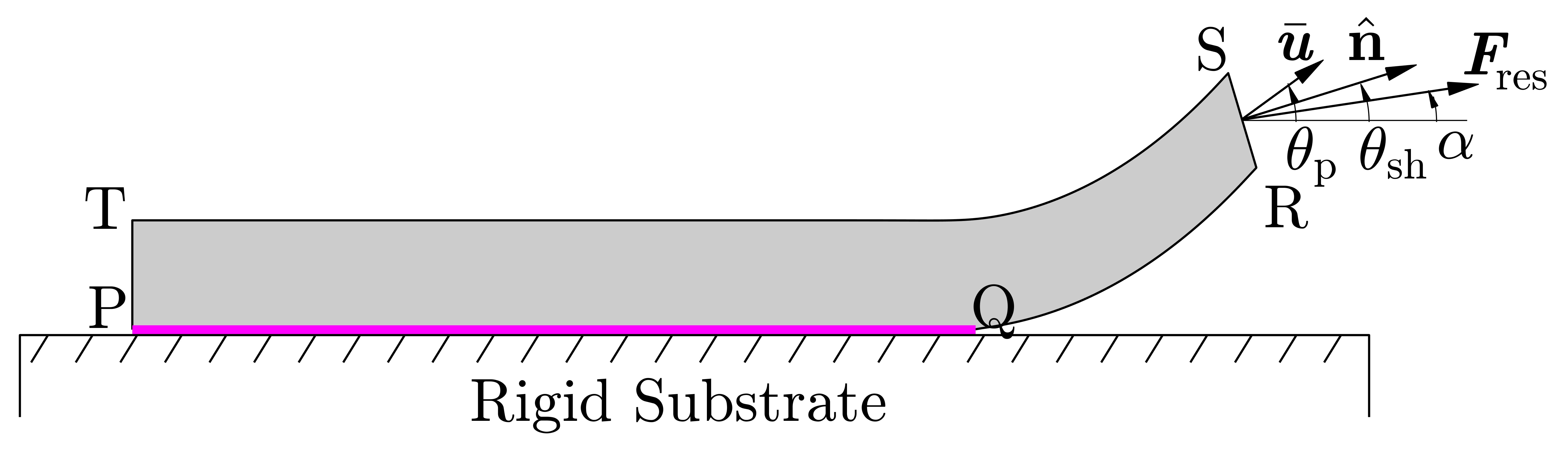}
	    		\caption{Peeling from a rotated configuration. Here, $\mathbf{\hat{n}}$ denotes the normal to the cross-section and $\theta_\mathrm{sh}$ is the angle $\mathbf{\hat{n}}$ makes with the horizontal, while $\alpha$ is the angle resultant force $\boldsymbol{F}_\mathrm{res}$ makes with the horizontal.}
	    		\label{fig:strip_rot}
	    	\end{center}
	    \end{figure}
	    
	    The Lennard-Jones (L-J) potential is considered in order to describe van der Waals interactions between the spatula and the rigid substrate. For any two molecules separated by a distance $d$ the L-J potential $\phi$, in addition to being a function of $d$, is characterized by an energy scale $\epsilon$ and length scale $d_o$. It is given by \cite{Israelachvili_book}
	    \begin{equation}\label{eq:Lennard_Jones}
	    \phi(d)  := \epsilon \left(\dfrac{{d}_o}{{d}} \right)^{12} - \, 2\,\epsilon \left(\dfrac{{d}_o}{{d}} \right)^6 \,.
	    \end{equation} 
    	
	    Due to these van der Waals interactions, a force is needed to peel the strip off the rigid substrate. The corresponding adhesive contact tractions acting on the spatula during peeling are obtained using the coarse-grained contact model of Sauer and Li \cite{RogerLi2007}. The total potential corresponding to all the interacting molecules of the spatula and the substrate is obtained by summing the individual interactions $\phi(d)$. Then, the gradient of this total interaction potential gives the adhesive contact traction as (see Sauer and Wriggers \cite{Sauer2009a} for a detailed derivation)
	    \begin{equation}
	    \label{eq:Adhesive_Traction_Ref}
	    \boldsymbol{T}\!_\mathrm{a}(d_\mathrm{s},\boldsymbol{n}_\mathrm{p}) = \frac{{A}_\mathrm{H}}{2\pi d_o^3}\,\left[  \frac{1}{45} \left(\frac{d_o}{d_\mathrm{s}}\right)^9 - \;\; \frac{1}{3} \left(\frac{d_o}{d_\mathrm{s}}\right)^3  \right]\,\boldsymbol{n}_\mathrm{p}\,,
	    \end{equation}
	    where $A_\mathrm{H}$ is the Hamaker's constant, $d_\mathrm{s}$ is the minimum distance between the spatula and the substrate and $\boldsymbol{n}_\mathrm{p}$ denotes the surface normal to the substrate. 
	    
	    The friction forces during adhesion are obtained following ``Model EA" of Mergel et al. \cite{Mergel2018}. Similar to the Coulomb friction model, the sticking and sliding phases are defined based on a threshold value $T_\mathrm{sl}$ for the magnitude of the tangential (or frictional) traction $\boldsymbol{T}\!_\mathrm{t}$. The sliding threshold, which is a function of $d_\mathrm{s}$, is dependent on the magnitude of the adhesive traction $T\!_\mathrm{a} = \left\Vert\boldsymbol{T}\!_\mathrm{a}\right\Vert$ and the friction coefficient $\mu_\mathrm{s}$, and is defined as
	    \begin{equation}
	    T_\mathrm{sl}(d_\mathrm{s}) = \begin{cases} 
	    \displaystyle\frac{\mu^{}_{\mathrm{s}}}{J_{\text{cl}}}\Big[T\!_\mathrm{a}(d_\mathrm{s}) - T\!_\mathrm{a}(d_\mathrm{cut})\Big], &  \quad d_\mathrm{s} < d_{\text{cut}}, \\
	    \displaystyle \quad \quad \quad 0, & \quad d_\mathrm{s} \geq d_{\text{cut}},
	    \end{cases} 
	    \label{eq:Sliding_Threshold}
	    \end{equation}
	    where $J_\mathrm{cl}$ is the local surface stretch of the substrate and is equal to unity if the substrate is rigid. The cut-off distance $d_\mathrm{cut}$ is the distance beyond which there is no frictional force. It is defined in terms of the equilibrium distance $d_\mathrm{eq}$, where the adhesive traction is zero, and the distance $d_\mathrm{max}$, at which the adhesive traction magnitude obtains the global minimum (maximum attraction),
	    \begin{equation}
	    d_\mathrm{cut} := s_\mathrm{cut}\,d_{\text{max}} + (1-s_\mathrm{cut})\,d_\mathrm{eq}, \quad \quad s_\mathrm{cut} \; \in \; [0,1]\,.
	    \end{equation}
	    
	    Following the experimental observations that the kinetic and static friction forces are comparable for biological adhesives \cite{Mergel2018}, the tangential contact traction $\boldsymbol{T}\!_\mathrm{t}$ can then be obtained as 
	    \begin{equation}
	    \label{eq:friction_traction}
	    \left\Vert\boldsymbol{T}\!_\mathrm{t} \right\Vert\begin{cases} 
	    = T_\mathrm{sl}  &  \quad \text{for sliding,} \\
	    < T_\mathrm{sl} & \quad \text{for sticking,}
	    \end{cases} 
	    \end{equation}
	    which is evaluated computationally using a predictor-corrector algorithm as discussed in \cite{Gouravaraju2020}. The total contact traction is then given by $\boldsymbol{T}\!_\mathrm{c} = \boldsymbol{T}\!_\mathrm{a} + \boldsymbol{T}\!_\mathrm{t}$.
	    
	    The elastic response of the spatula is modeled using a Neo-Hookean material model for which the strain energy density function is given as \cite{bonet2008}
	    \begin{equation}
	    \Psi = \frac{\mu}{2}\Big(\mathrm{tr}(\boldsymbol{b})-3\Big)\, + \, \frac{\lambda}{2}\left(\ln J\right)^2 \, - \, \mu\ln J, \label{eq:strain_energy_fun}
	    \end{equation}
	    where $\lambda$ and $\mu$ are Lam\'{e} constants, $\boldsymbol{b} = \boldsymbol{F}\boldsymbol{F}^\mathrm{T}$ is the left Cauchy-Green deformation tensor, and $J$ is the determinant of the deformation gradient $\boldsymbol{F}$. The strain energy $\Pi_\mathrm{int}$ is then given by
	    \begin{equation}
	    \Pi_\mathrm{int} = \int_{\mathcal{B}_0} \,\Psi \, \mathrm{d}V\,.
	    \end{equation}
	    
 		The absolute value of the adhesion energy required to completely separate the spatula from its initial configuration shown in Figure~{\ref{fig:strip_orig}} is obtained from the work of adhesion $w_\mathrm{a}$ as \cite{Sauer2013}
	    \begin{equation}
	    \Pi_{\mathrm{a},0} = \mathrm{A}_\mathrm{pad}\, w_\mathrm{a} \quad \mathrm{where} \quad w_\mathrm{a} = - \int_{d_\mathrm{eq}}^\infty \, \Vert \boldsymbol{T}_\mathrm{a}(d_\mathrm{s}) \Vert \, \mathrm{d}d_\mathrm{s} = \sqrt[3]{15} \, \dfrac{A_\mathrm{H}}{16 \pi d_o^2}\,,	
	    \end{equation}
	    and where $A_\mathrm{pad}$ is the initial area of the spatula pad in contact with the substrate. So, as the spatula is gradually peeled off the substrate the adhesion energy $\Pi_\mathrm{a}$ increases from $-\Pi_{\mathrm{a},0}$ and eventually becomes zero when the spatula is completely peeled off.
	    
	   \section{Application of boundary conditions} \label{sec:peel_appl}
	    Three different types of peeling simulations are carried out here:
	    \begin{enumerate}
	    	\item[1.] In ``Type I" simulations, the spatula is peeled off from its initial configuration by applying a displacement $\bar{\boldsymbol{u}}$ to the right end (RS) at an angle called peeling angle $\theta_\mathrm{p}$ as shown in Figure~\ref{fig:strip_orig}.
	    	
	    	\item[2.] In ``Type II" simulations, first an external rotation is applied to the right end (RS) of the strip. After, achieving a desired rotation angle $\theta_\mathrm{sh}$ (called the shaft angle) on the right end, a displacement $\bar{\boldsymbol{u}}$ is applied at the constant peeling angle $\theta_\mathrm{p} = 90^\circ$ (see Figure~\ref{fig:strip_rot}).
	    	
	    	\item[3.] ``Type III" simulations denote the special case $\theta_\mathrm{p} = \theta_\mathrm{sh}$ in ``Type II" simulations.  (Figure~\ref{fig:strip_rot}).
	    \end{enumerate}
    	
	    In all the simulations, the displacement $\bar{\boldsymbol{u}}$ is applied through its components $u_x = \bar{u}\cos\theta_\mathrm{p}$ and $u_y = \bar{u}\sin\theta_\mathrm{p}$, which results in a tangentially constrained motion of the spatula shaft. As a consequence, the resultant pull-off froce $\boldsymbol{F}\!_\mathrm{res}$ is not parallel to the applied displacement $\bar{\boldsymbol{u}}$. Instead $\boldsymbol{F}\!_\mathrm{res}$ acts in the direction (see Figure~\ref{fig:strip_rot})
	    \begin{equation}
	    \alpha = \mathrm{arctan}(F\!_\mathrm{N}/F\!_\mathrm{T})\,,
	    \end{equation} 
	    where $F\!_\mathrm{N}$ and $F\!_\mathrm{T}$ are its normal and tangential components.
	    
	    A total of $240\times12$ finite elements are used along the $x$ and $y$ directions for the strip. The finite element model of Gouravaraju et al. \cite{Gouravaraju2020} based on the enriched contact disretization of Sauer \cite{RogerEnriched2011} is used to simulate the peeling. Plane strain\footnote[2]{As the spatula is very wide (up to a few hundred nm), plane strain is a reasonable simplification of the full 3D case.} simulations are carried out using the parameters listed in Table~\ref{tab:table1} \cite{Tian2006,Autumn2006d,RogerEnriched2011}. The initial area of the spatula pad is taken as $A_\mathrm{pad} = 49,524 R_0^2$ \cite{Sauer2013}. From the parameters in Table~\ref{tab:table1} and Figure~\ref{fig:strip_orig} the width of the pad for the strip configuration becomes $w_\mathrm{pad}=330.16R_0$.
	    
	    \begin{table}[h!]
	    	\begin{center}
	    		\caption{\label{tab:table1}%
	    			Geometrical, material, and adhesion parameters used in the current study. Here, $R_0 = 1$ nm is introduced for normalization.}
	    		\begin{tabular}{|c|c|}
	    			\hline
	    			& \\[-1em]
	    			\textrm{Length ($L$)}& $200R_0$ \\
	    			& \\[-1em]
	    			\hline
	    			& \\[-1em]
	    			\textrm{Height ($h$)}& $10R_0$ \\
	    			& \\[-1em]
	    			\hline
	    			& \\[-1em]
	    			\textrm{Young's Modulus ($E_0$)}& $2$ GPa \\
	    			& \\[-1em]
	    			\hline
	    			& \\[-1em]
	    			\textrm{Poisson's ratio ($\nu$)}& 0.2 \\
	    			&\\[-1em]
	    			\hline
	    			& \\[-1em]
	    			\textrm{Friction coefficient ($\mu_\mathrm{s}$)}& 0.3\\
	    			& \\[-1em]
	    			\hline
	    			& \\[-1em]
	    			\textrm{Equilibrium distance ($d_o$)}& 0.4 nm\\
	    			& \\[-1em]
	    			\hline
	    			&\\[-1em]
	    			\textrm{Hamaker's constant ($A_\mathrm{H}$)}& $10^{-19}$ J\\
	    			\hline
	    		\end{tabular}
	    	\end{center}
	    \end{table}

	    \section{Results and Discussion}
	    In this section, the numerical results obtained from the finite element simulations are presented, and their significance in explaining the critical detachment angle at the spatula level is discussed.
	    
	    \subsection{Pull-off forces} \label{sec:pull_forces}
	    \begin{figure}[h!]
	    	\begin{center}
	    		\includegraphics[scale=0.31]{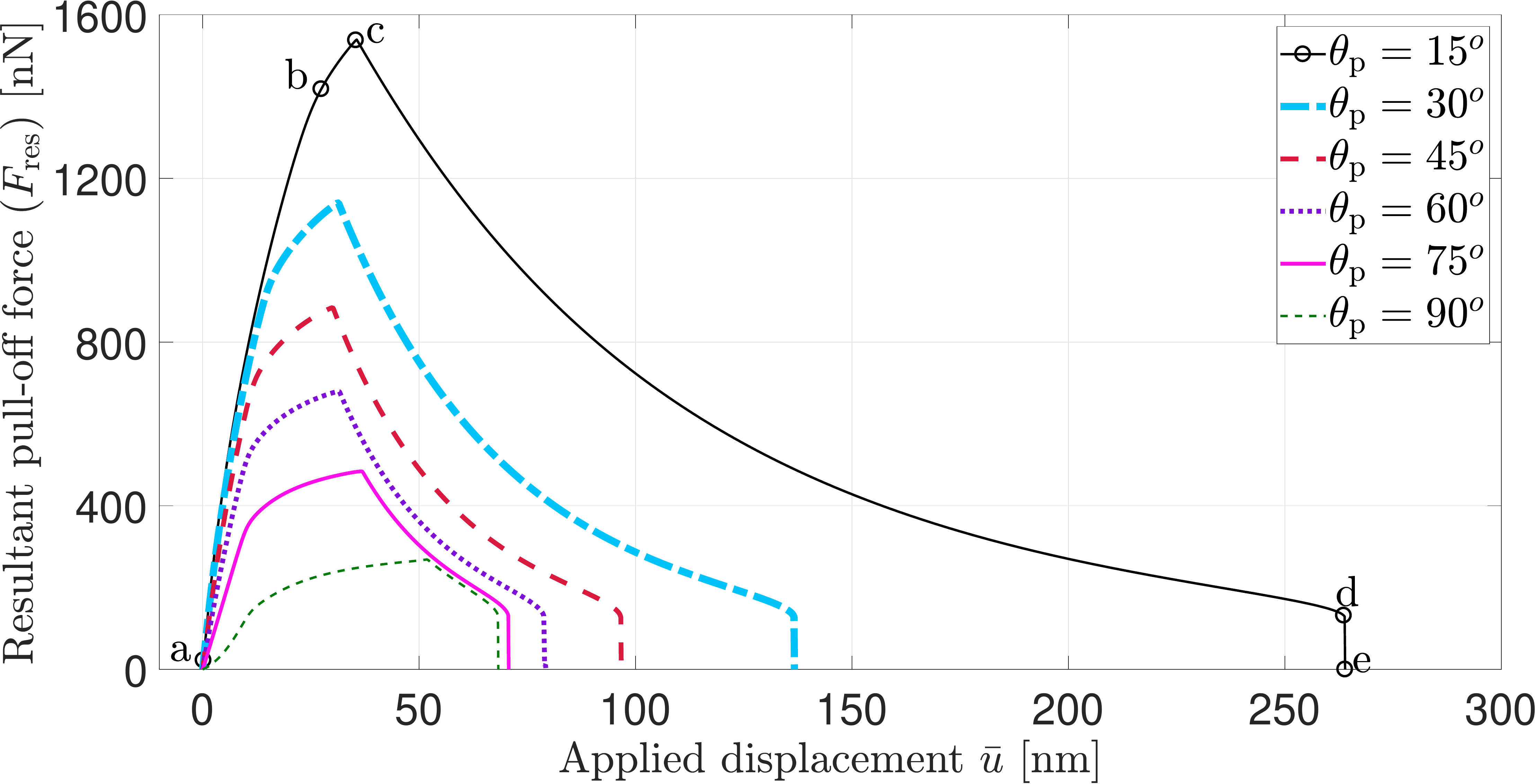} 			   
	    		\caption{Evolution of the resultant pull-off force $F\!_\mathrm{res} = \Vert \boldsymbol{F}\!_\mathrm{res}\Vert$ with the applied displacement $\bar{u} = \Vert \bar{\boldsymbol{u}} \Vert$ for different peeling angles $\theta_\mathrm{p}$ (``Type I" simulations).}   	 \label{fig:res_f}	
	    	\end{center}
	    \end{figure}
\begin{figure}[h!]
    	\begin{center}
    		\includegraphics[scale=0.31]{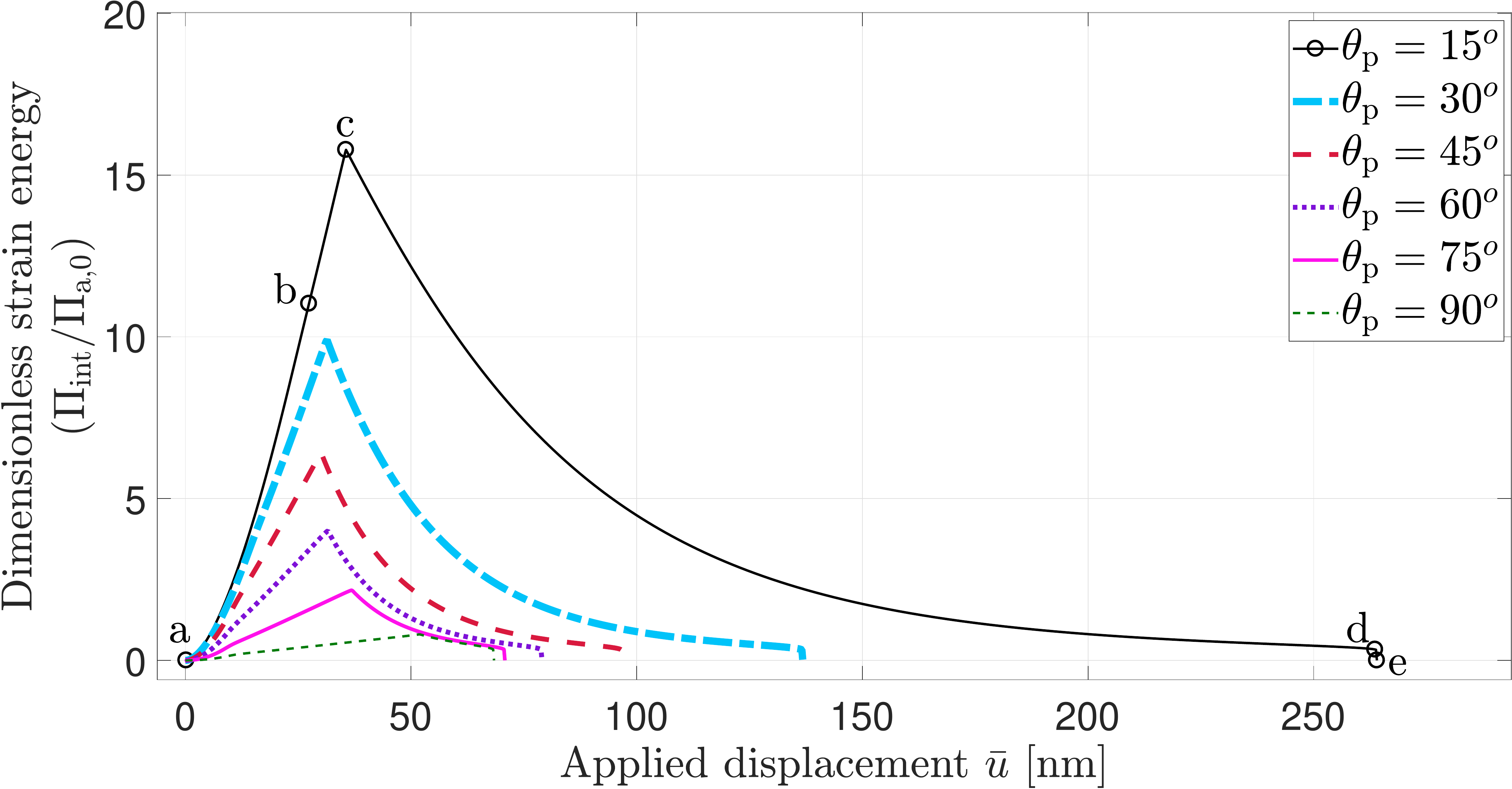} 			   
    		\caption{Evolution of the dimensionless strain energy with the applied displacement $\bar{u}= \Vert \bar{\boldsymbol{u}} \Vert$ for different peeling angles $\theta_\mathrm{p}$ (``Type I" simulations). Here, $\Pi_{\mathrm{a},0} = 1.523 \times 10^{-15}$ J.}   	 \label{fig:strain_energy}	
    	\end{center}
    	\end{figure}
\begin{figure}[h!]
    	\begin{center}
    		\includegraphics[scale=0.31]{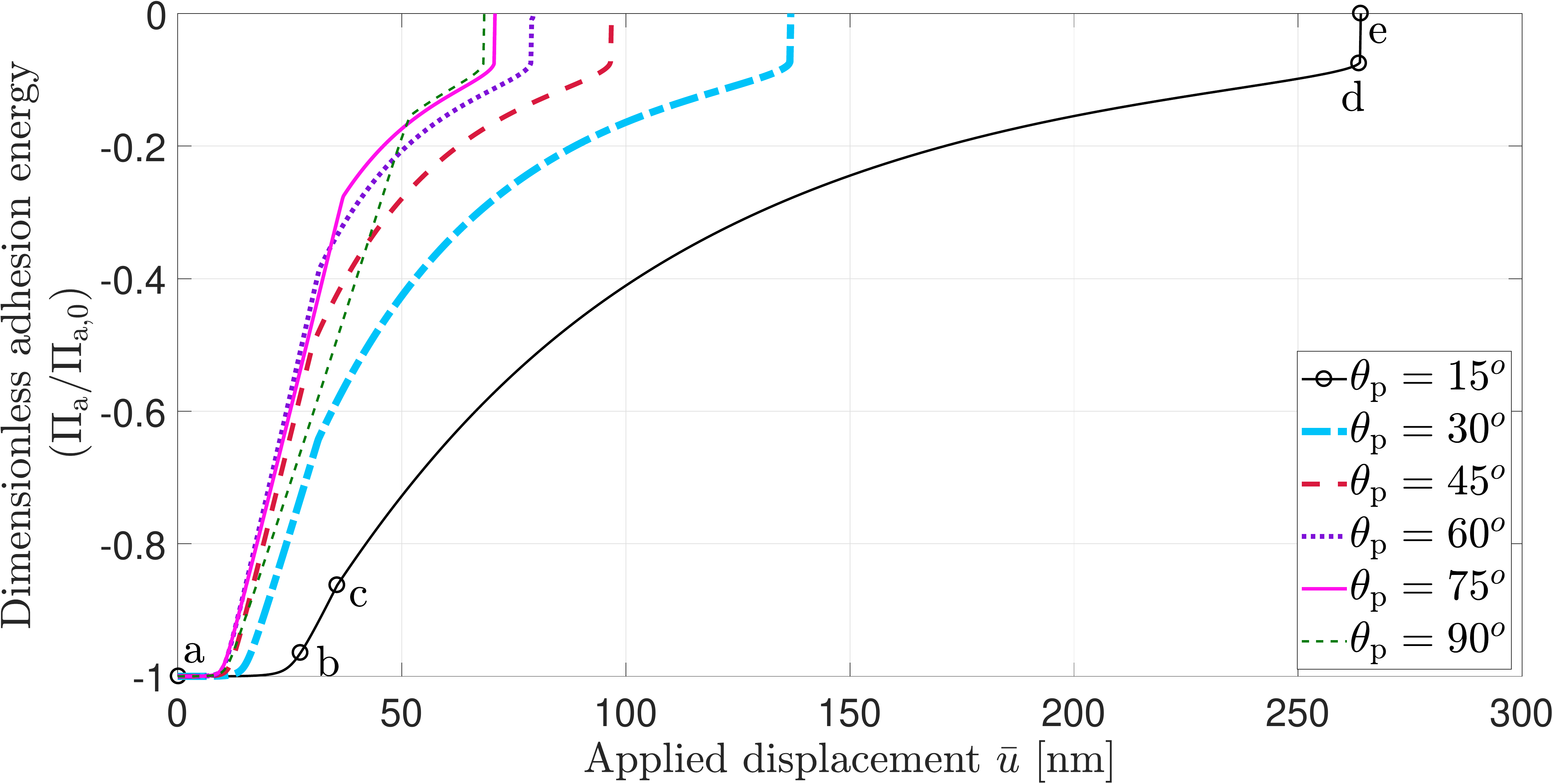} 			   
    		\caption{Evolution of the dimensionless adhesion energy with the applied displacement $\bar{u}= \Vert \bar{\boldsymbol{u}} \Vert$ for different peeling angles $\theta_\mathrm{p}$ (``Type I" simulations). Here, $\Pi_{\mathrm{a},0} = 1.523 \times 10^{-15}$ J.}   	 \label{fig:adh_energy}	
    	\end{center}
   	    \end{figure}

	    It has been recently shown by Gouravaraju et al. \cite{Gouravaraju2020} that as the spatula is peeled off the substrate, it goes through two major phases. This is illustrated with the resultant pull-off force versus applied displacement curve shown in Figure~\ref{fig:res_f}. In the initial phase (marked ``a" to ``c" for the case of $\theta_\mathrm{p} = 15^\circ$), the spatula undergoes stretching due to partial sliding near the peeling front and thus accumulates strain energy (see Figure~\ref{fig:strain_energy}). As a result, the pull-off force also increases. Point ``b" marks the first instance at which the spatula pad starts to peel off the substrate. As a consequence the adhesion energy changes from its initial value $\Pi_{\mathrm{a},0}$ as shown in Figure~{\ref{fig:adh_energy}}. After reaching a maximum, in the second phase (``c" to ``d" in Figure~\ref{fig:res_f}), the spatula relaxes as full sliding ensues, gradually releasing the stored energy. After reaching a critical point ``d", further application of the displacement results in a sudden release of the remaining stored energy and complete peel-off of the spatula pad from the substrate at point ``e". Numerical simulations for various peeling angles resulted in similar peeling behavior (see Figures~\ref{fig:strain_energy} and \ref{fig:adh_energy}). Further, it is observed that the tangential friction force is the major contributor to the resultant pull-off force.
	    
	    \subsection{Critical detachment angle}\label{sec:fric_adhes}		
	    Autumn et al. \cite{Autumn2006b} have observed that irrespective of the applied load, at each level in the hierarchy of the gecko adhesive system down to the setae, the structures detaches from the substrate when the angle between the resultant force vector and the substrate $\alpha$, equals the critical detachment angle $\alpha^*$. This critical detachment angle varies among setae ($\alpha^*_{\mathrm{seta}} = 30^\circ$), seta arrays ($\alpha^*_{\mathrm{array}} = 24.6\pm0.9^\circ$), and toes ($\alpha^*_{\mathrm{toe}} = 25.5\pm0.2^\circ$). However, since no experiments have been performed for the spatula, it is not clear if similar values are found at the spatula level.
	    
	    \begin{figure}[h!]
	    	\begin{center}
	    		\includegraphics[scale=0.31]{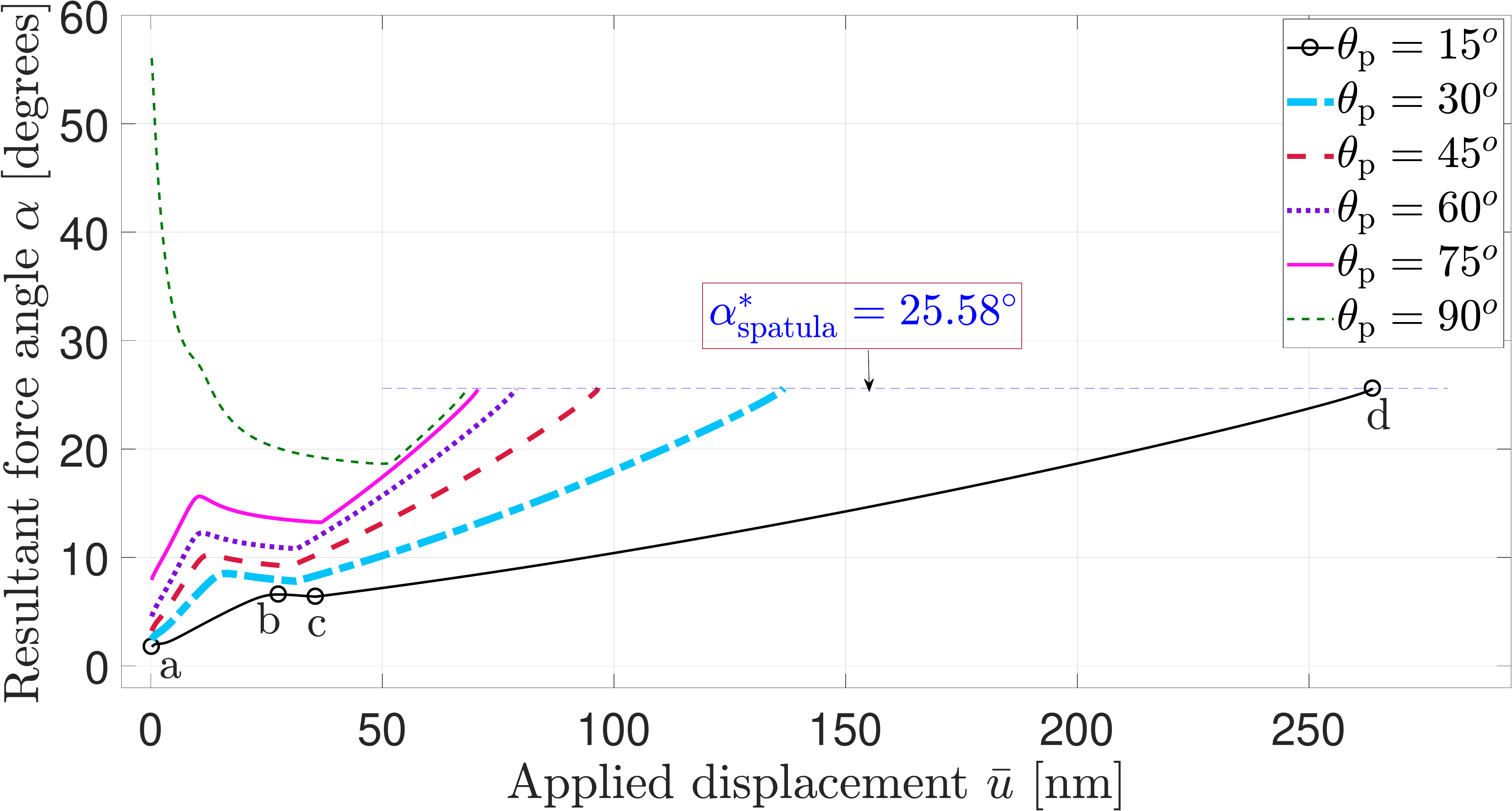} 			   
	    		\scriptsize{\caption{Evolution of the resultant force angle $\alpha$ with the applied displacement $\bar{u}= \Vert \bar{\boldsymbol{u}} \Vert$ for different peeling angles $\theta_\mathrm{p}$ (``Type I" simulations). At detachment (point ``d") the common value $\alpha^\ast_\mathrm{spatula} = 25.58^\circ$ is observed.}   	 \label{fig:Angle_Res_Forc}}	
	    	\end{center}
	    \end{figure}
	    
\begin{figure}[h!]
	    	\begin{center}
	    		\includegraphics[scale=0.31]{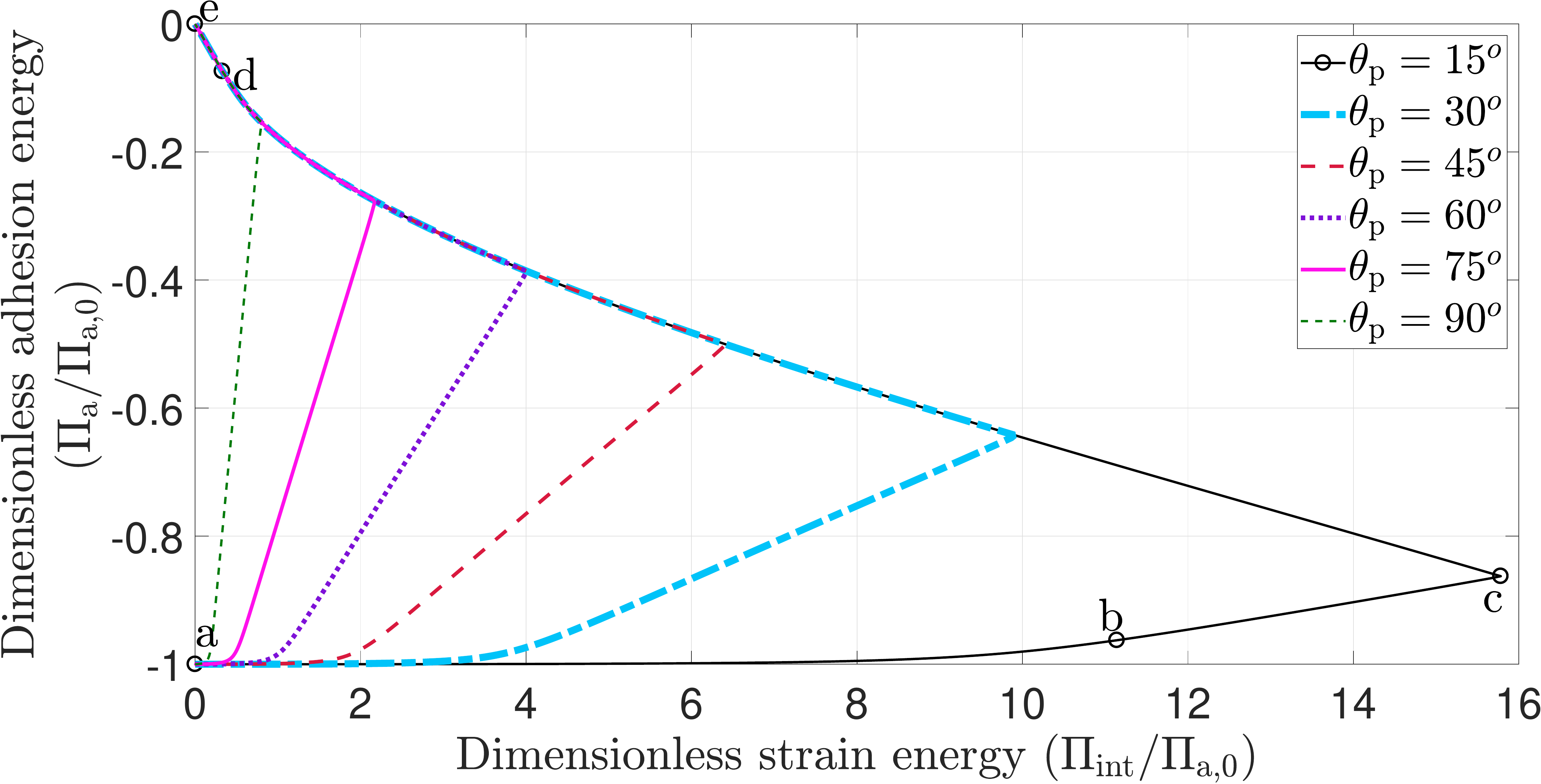} 			
	    		\caption{Evolution of the adhesion energy with the strain energy for different peeling angles $\theta_\mathrm{p}$ (``Type I" simulations). Here, $\Pi_{\mathrm{a},0} = 1.523 \times 10^{-15}$ J.}   	 \label{fig:SEvAE}
	    	\end{center}
	    \end{figure}

	    Figure~{\ref{fig:Angle_Res_Forc}} shows the resultant force angle $\alpha$ with the applied displacement $\bar{u}$ for different peeling angles $\theta_\mathrm{p}$. It can be seen that the resultant force angle changes throughout the peeling process. Except for $\theta_\mathrm{p} = 90^\circ$, the resultant force angle curves follow similar paths. This is illustrated with the help of the curve for $\theta_\mathrm{p} = 15^\circ$. The points ``a" to ``d" on this curve directly correspond to those in Figures~{\ref{fig:res_f}}, {\ref{fig:strain_energy}} and {\ref{fig:adh_energy}}. The point ``e" is not shown here as at this point both the normal and tangential forces become zero. The resultant force angle $\alpha$ initially increases up to a point ``b", which is when the pad (PQ in Figure~\ref{fig:strip_orig}) starts to detach from the substrate. The resultant force angle then decreases until it reaches the point ``c", which, as seen from Figures~{\ref{fig:res_f}} and {\ref{fig:strain_energy}}, is the point where the strain energy and the resultant pull-off force reach their maximum values. Beyond the point ``c", $\alpha$ increases monotonically until the critical point ``d" which is the jump-off contact point. For the case of $\theta_\mathrm{p} = 90^\circ$, points ``a" and ``b" coincide as the spatula pad starts to peel off as soon as the displacement is applied. This can be clearly seen from the evolution of the dimensionless strain energy and the dimensionless adhesion energy for different peeling angles shown in Figure~{\ref{fig:SEvAE}}. For $\theta_\mathrm{p}=90^\circ$, there is a sharp increase in the adhesion energy from its initial value $-\Pi_{\mathrm{a},0}$. 
	    
	    The most important observation from Figure~\ref{fig:Angle_Res_Forc} is that irrespective of the peeling angle $\theta_\mathrm{p}$, the spatula detaches from the substrate at a constant detachment angle of $\alpha^*_{\mathrm{spatula}} = 25.58\pm0.07^\circ$, even though $\alpha$ changes throughout the peeling process. 
	    
	    \begin{figure}[h!]
	    	\begin{center}
	    		\includegraphics[scale=0.31]{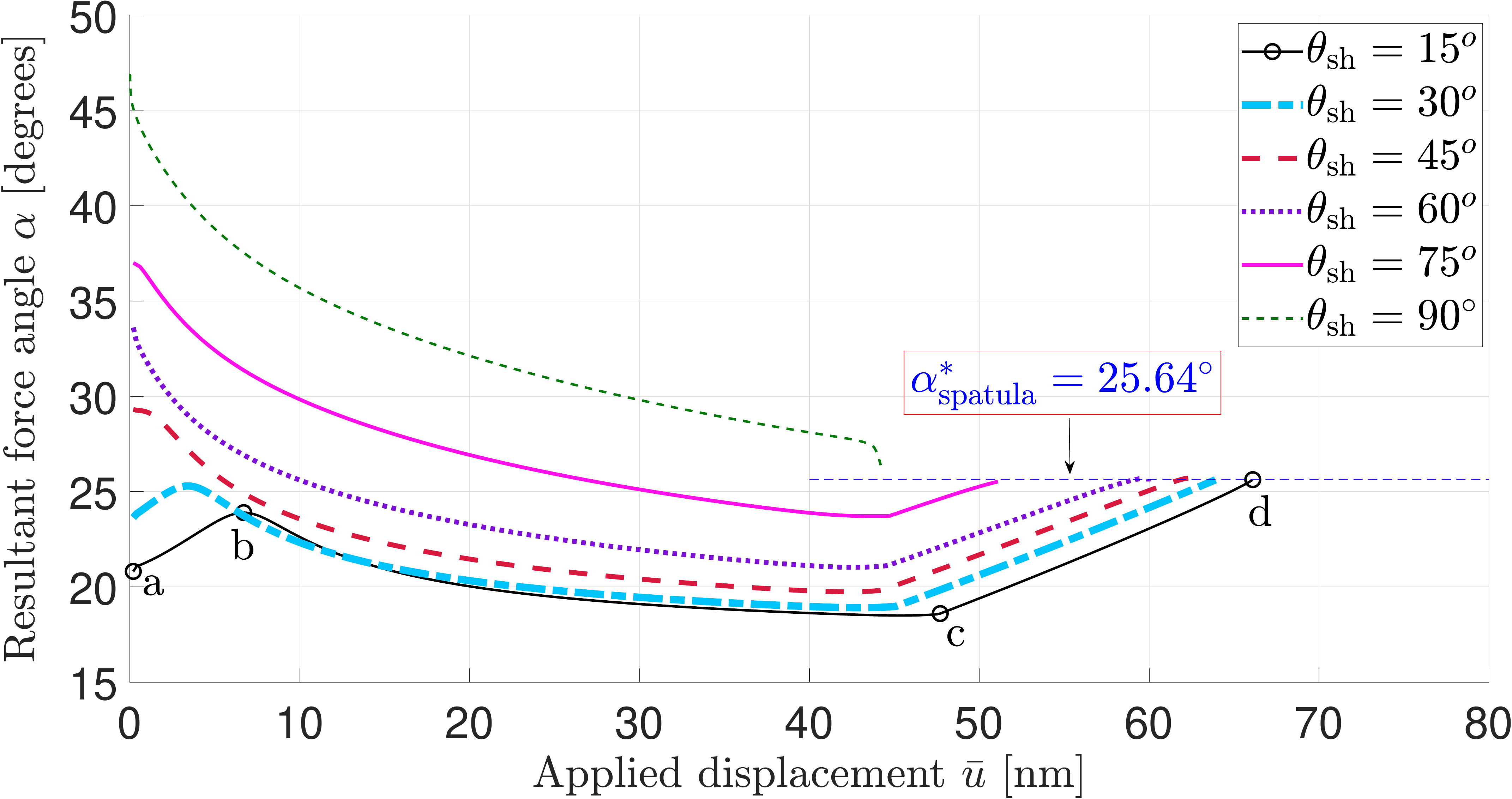} 			\caption{Evolution of the resultant force angle $\alpha$ with the applied displacement $\bar{u}= \Vert \bar{\boldsymbol{u}} \Vert$ for different shaft angles $\theta_\mathrm{sh}$ and the peeling angle $\theta_\mathrm{p}=90^\circ$ (``Type II" simulations). At detachment (point ``d") the common value $\alpha^\ast_\mathrm{spatula} = 25.64^\circ$ is observed. }   	 \label{fig:alpha_diff_sh}
	    	\end{center}
	    \end{figure}
	    
	    \begin{figure}[]
	    	\begin{center}
	    		\includegraphics[scale=0.31]{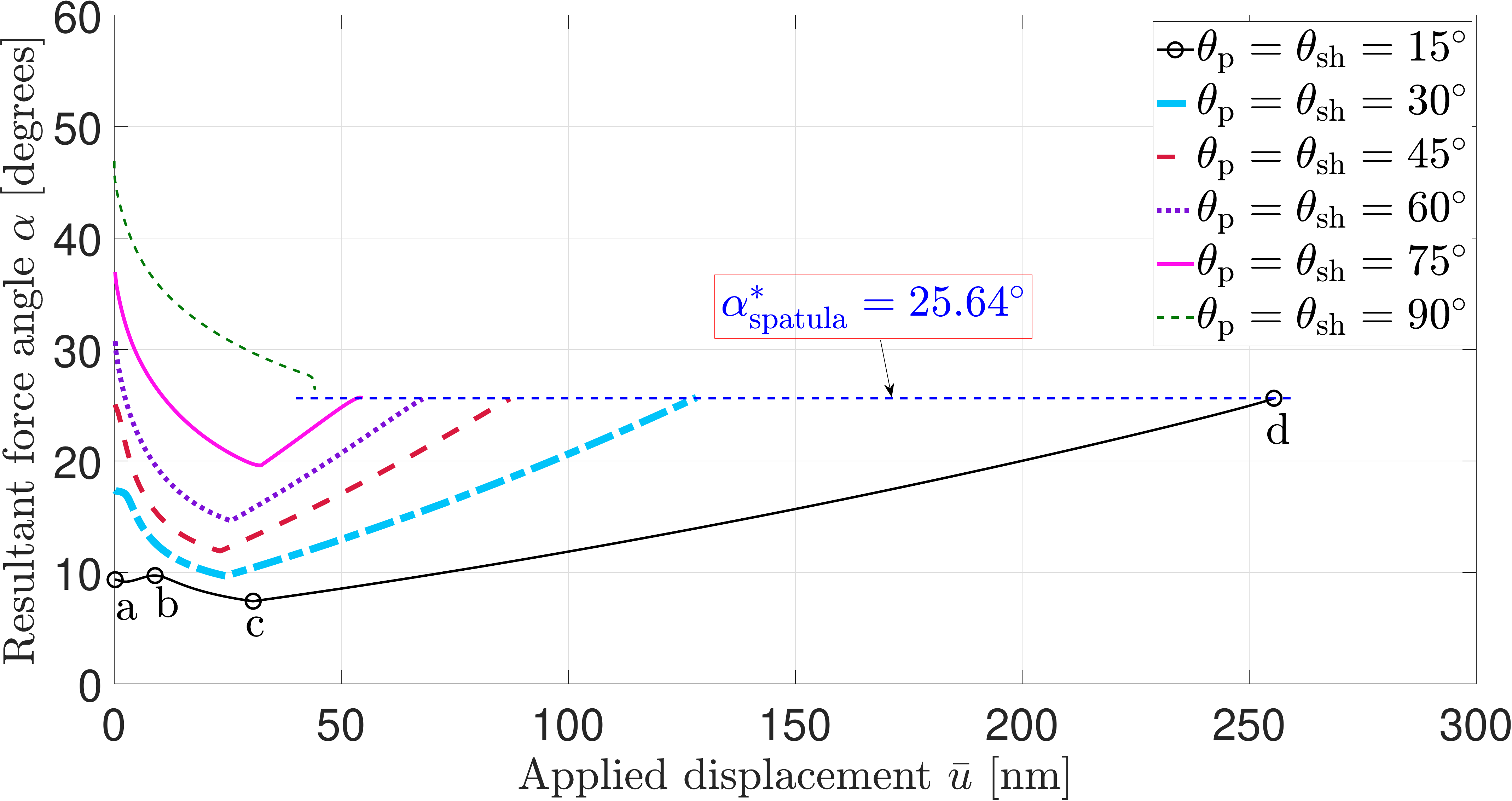} 			\caption{Evolution of the resultant force angle $\alpha$ with the applied displacement $\bar{u}= \Vert \bar{\boldsymbol{u}} \Vert$ for different peeling and shaft angles $\theta_\mathrm{p}=\theta_\mathrm{sh}$ (``Type III" simulations). At detachment (point ``d") the common value $\alpha^\ast_\mathrm{spatula} = 25.64^\circ$ is observed.}   	 \label{fig:alpha_sh_thetap}
	    	\end{center}
	    \end{figure}
    
	    Similar behavior is observed for ``Type II" simulations when the spatula is pulled at a constant peeling angle $\theta_\mathrm{p}=90^\circ$ from various pre-rotated configurations, i.e., with different shaft angles $\theta_\mathrm{sh}$ as shown in Figure~\ref{fig:alpha_diff_sh}. It can be seen that irrespective of the shaft angle the detachment angle is approximately the same for all peeling simulations and is equal to~$\alpha^*_\mathrm{spatula} = 25.58^\circ$. Furthermore, ``Type III" simulations (see Figure~\ref{fig:alpha_sh_thetap}) also show similar behavior with $\alpha^*_\mathrm{spatula} = 25.64^\circ$. From all these results, it is clear that irrespective of the spatula shaft angle and the peeling angle, the critical detachment angle remains nearly invariant. It should be noted that the points ``a" to ``d" marked in Figures~\ref{fig:alpha_diff_sh} and \ref{fig:alpha_sh_thetap} correspond to the instances discussed in section~\ref{sec:pull_forces} and shown in Figures~\ref{fig:res_f} to \ref{fig:adh_energy}. Table~\ref{tab:Det_Ang} lists the critical detachment angle for all three types of simulations. 
	    
		\begin{table}[ht]
			\begin{center}
				\caption{\label{tab:Det_Ang}%
					Critical detachment angle $\alpha^*_\mathrm{spatula}$ for different types of simulations.}
				\begin{tabular}{|p{2cm}|p{2cm}|p{2cm}|p{2cm}|p{2.4cm}|p{2cm}|}
					\hline
					\multicolumn{2}{|c|}{Type I ($\theta_\mathrm{sh}=0^\circ$)} &\multicolumn{2}{c|}{Type II ($\theta_\mathrm{p} = 90^\circ$)} & \multicolumn{2}{c|}{Type III ($\theta_\mathrm{p} = \theta_\mathrm{sh}$)}\\[1ex]
					\hline
					Peeling angle ($\theta_\mathrm{p}$) & Critical detachment angle ($\alpha^*_\mathrm{spatula}$) & Shaft angle ($\theta_\mathrm{sh}$)& Critical detachment angle ($\alpha^*_\mathrm{spatula}$)& Peeling angle ($\theta_\mathrm{p} = \theta_\mathrm{sh}$)& Critical detachment angle ($\alpha^*_\mathrm{spatula}$)\\
					\hline
					$10^\circ$ & $25.65^\circ$ & $10^\circ$ & $25.45^\circ$ & $10^\circ$ & $25.57^\circ$\\
					\hline 
					$15^\circ$ & $25.58^\circ$& $15^\circ$ & $25.63^\circ$& $15^\circ$ & $25.64^\circ$ \\
					\hline 					
					$20^\circ$ & $25.56^\circ$& $20^\circ$&$25.53^\circ$& $20^\circ$ &$25.41^\circ$\\
					\hline
					$25^\circ$ & $25.60^\circ$ &$25^\circ$&$25.48^\circ$ &$25^\circ$ &$25.42^\circ$ \\
					\hline
					$30^\circ$ & $25.61^\circ$ & $30^\circ$ &$25.62^\circ$ &$30^\circ$& $25.69^\circ$ \\
					\hline
					$35^\circ$ & $25.55^\circ$& $35^\circ$&$25.49^\circ$ & $35^\circ$& $25.46^\circ$\\
					\hline
					$40^\circ$ & $25.56^\circ$&$40^\circ$&$25.52^\circ$ &$40^\circ$ & $25.49^\circ$\\
					\hline
					$45^\circ$ & $25.62^\circ$&$45^\circ$&$25.70^\circ$ &$45^\circ$ & $25.57^\circ$  \\
					\hline
					$50^\circ$ & $25.65^\circ$&$50^\circ$&$25.55^\circ$ &$50^\circ$ & $25.67^\circ$\\
					\hline
					$55^\circ$ & $25.62^\circ$&$55^\circ$&$25.61^\circ$ &$55^\circ$ & $25.69^\circ$\\
					\hline
					$60^\circ$ & $25.54^\circ$&$60^\circ$&$25.69^\circ$ &$60^\circ$ & $25.68^\circ$\\
					\hline
					$65^\circ$ & $25.66^\circ$&$65^\circ$&$25.49^\circ$ &$65^\circ$ & $25.66^\circ$\\
					\hline
					$70^\circ$ & $25.52^\circ$&$70^\circ$&$25.63^\circ$ &$70^\circ$ & $25.70^\circ$\\
					\hline
					$75^\circ$ & $25.50^\circ$&$75^\circ$&$25.53^\circ$ &$75^\circ$ & $25.50^\circ$\\
					\hline
					$80^\circ$ & $25.70^\circ$&$80^\circ$&$25.48^\circ$ &$80^\circ$& $25.65^\circ$\\
					\hline
					$85^\circ$ & $25.45^\circ$&$85^\circ$&$25.63^\circ$ &$85^\circ$&$25.69^\circ$\\
					\hline
					$90^\circ$ & $25.50^\circ$&$90^\circ$&$26.39^\circ$ &$90^\circ$ & $26.39^\circ$\\
					\hline
				\end{tabular}
			\end{center}
		\end{table}
	    
	    \begin{figure}[h!]
	    	\begin{center}
	    		\includegraphics[scale=0.31]{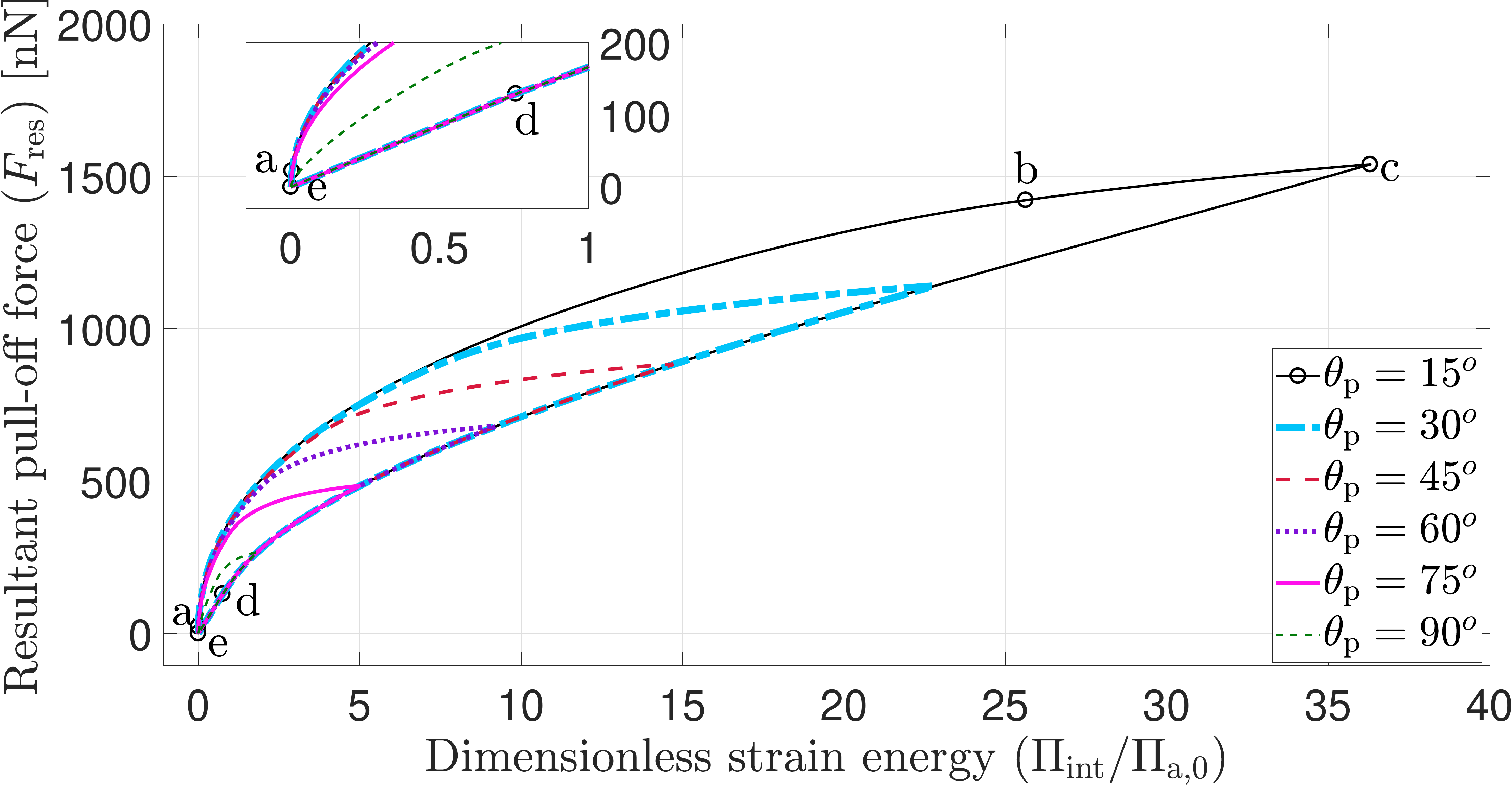} 			\caption{Resultant pull-off force and strain energy evolution for different peeling angles $\theta_\mathrm{p}$ for ``Type I" simulations. Here, $\Pi_{\mathrm{a},0} = 1.523 \times 10^{-15}$ J. The inset shows an enlargement of the curves at the jump-off contact point ``d".}   	 \label{fig:FvSE}
	    	\end{center}
	    \end{figure}
    
    	To understand the invariance of the detachment angle, the evolution of the resultant pull-off force vs. the strain energy up to the critical point at which spatula snap-off occurs (point ``d") is plotted for different peeling angles $\theta_\mathrm{p}$, see Figure~\ref{fig:FvSE}. These curves show the strain energy and pull-off force values for which the spatula maintains attachment as it is peeled off the substrate. Even as the sliding starts (point ``c") and the contact becomes unstable, the spatula resists detachment until it reaches the critical point ``d".  Also, the curves follow the same paths for the initial stretching and also after the sliding starts. Most importantly, irrespective of the peeling angle and the maximum force reached during the peeling process, the spatula reaches approximately the same critical point ``d" (see inset in Figure~\ref{fig:FvSE}). These curves are similar to the \emph{stability envelopes} for the peeling of thin tapes shown by Federle and Labonte \cite{Labonte2019}. These stability envelopes are obtained by plotting the pull-off force as a function of the pre-strain. For small peeling angles, the tape stretches such that the pre-strain exceeds a critical value called the minimum pre-strain $\varepsilon_\mathrm{min}$. Federle and Labonte \cite{Labonte2019} have shown that once the tape is stretched beyond $\varepsilon_\mathrm{min}$, it can be spontaneously detached by decreasing the applied force below the minimum force required to stabilize the contact. In the current study, these minimum force values are given by the resultant pull-off force values on the curve from ``c" to ``d".  Further, the minimum strain of Federle and Labonte \cite{Labonte2019} can be related to the strain energy at point ``d".     	
    	
    	Similar behavior can also be observed in Figure~\ref{fig:SEvAE}. It can be seen that all the peeling curves for all the peeling angles follow similar paths. Moreover, it is observed that at the snap-off point ``d",  all the peeling curves reach approximately the same energy state (see inset in Figure~\ref{fig:SEvAE}). These results reveal an interesting observation that irrespective of the peeling angle, the spatula follows similar paths in detaching from the substrate and finally reaches the same critical energy state beyond which it cannot sustain any more loading and detaches from the substrate completely. 

	    According to the  ``frictional adhesion" model of Autumn et al. \cite{Autumn2006b}, the normal adhesive force is limited by the frictional force and the critical detachment angle $\alpha^*$ and is given by
	    \begin{equation}
	    \tan \alpha^* \geq \frac{F_\mathrm{N}}{F_\mathrm{T}}\,.
	    \label{eq:Fric_Adhesion}
	    \end{equation} 
	    In the current work, where $\alpha^*_{\mathrm{spatula}} \approx 25.6^\circ$, the maximum friction force must always be greater than $2.1$ times the maximum normal adhesive force according to Eq.~\ref{eq:Fric_Adhesion}. From, the resultant force angle curves shown in Figure~\ref{fig:Angle_Res_Forc}, it is clear that the ratio of the maximum friction force to the maximum normal force is always more than $2.1$. For $\theta_p = 10^\circ$, the maximum value of the friction force can be as high as $8.6$ times the maximum value of the normal force. As a consequence even when the spatula starts to fully slide on the substrate (point ``c"), the spatula is still attached to the substrate until the critical point ``d" is reached. However, for $\theta_\mathrm{p} = \theta_\mathrm{sh} = 90^\circ$ this is not true (see Figures~\ref{fig:alpha_diff_sh} and $\ref{fig:alpha_sh_thetap}$), and as soon as the force maximum is reached and the spatula pad starts to fully slide on the substrate, it snaps-off from the substrate, i.e. points ``c" and ``d" coincide for $\theta_\mathrm{p} = \theta_\mathrm{sh} = 90^\circ$. This supports the experimental observations of Autumn et al. \cite{Autumn2006c} that when geckos adhere to a substrate, they generate much greater forces than are required for them to stay attached to the substrate according to Eq.~(\ref{eq:Fric_Adhesion}) (a shear force of $5$ times the adhesive force for the gecko front legs was measured by Autumn et al. \cite{Autumn2006c}, while we find a shear force as high as $8.6$ times the adhesive force for the spatula). From these results, it can be concluded that adhesive friction, starting with the spatula level, is present at all hierarchy levels in the gecko adhesive system. Moreover, the critical detachment angle of $25.6^\circ$ implies that at detachment, the adhesive force is about half of the shear force. This particular trend is observed to be true for many of the climbing animals which use ``dry" as well as ``wet" adhesion \cite{Labonte2019}. This shows that the current model can be employed effectively to study other kinds of biological adhesive systems. 
	    
	    \subsection{Influence of spatula pad thickness on the critical detachment angle}
	     \begin{figure}[h!]
	    	\begin{center}
	    		\includegraphics[scale=0.3]{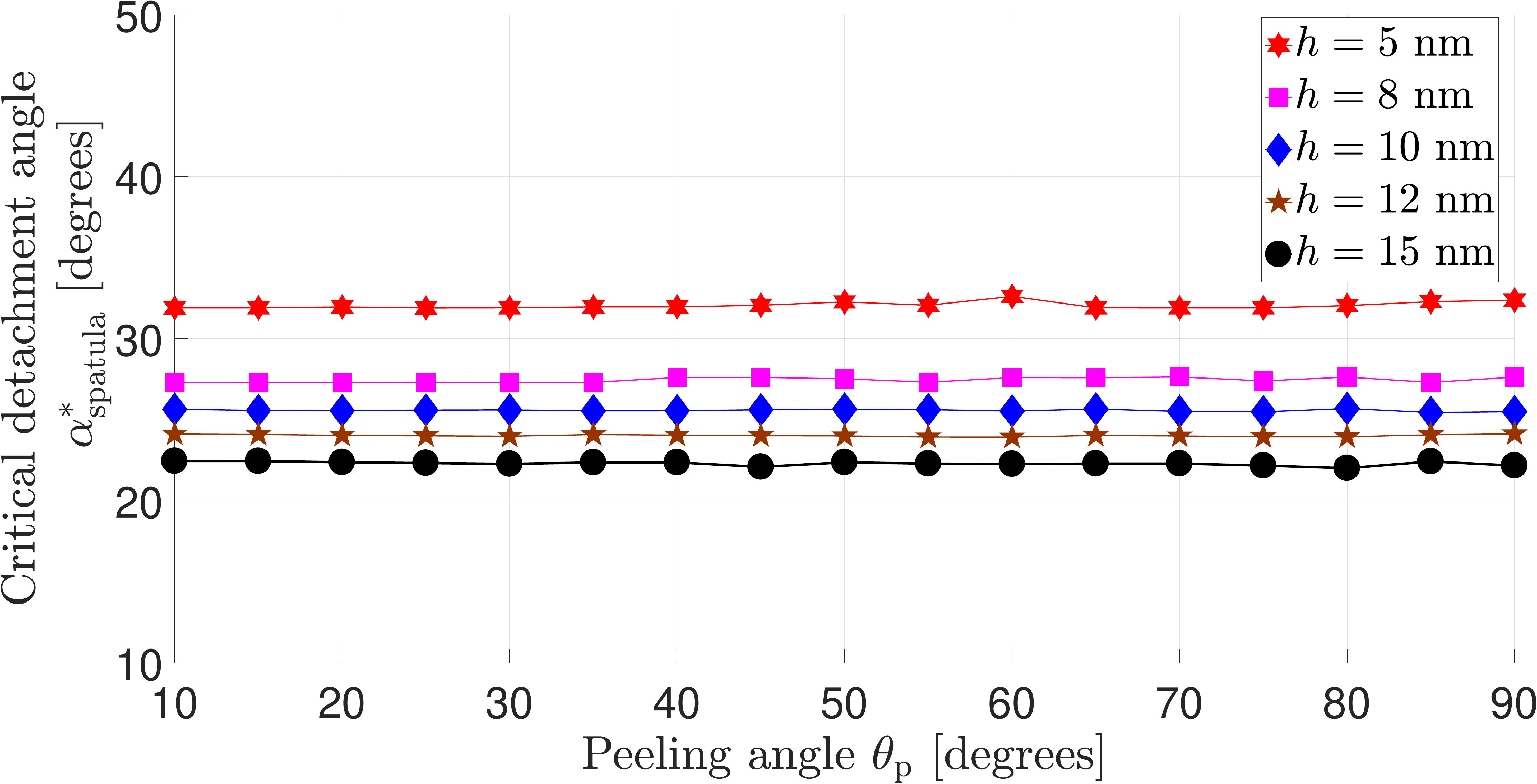} 	
	    		\caption{Variation of the critical detachment angle $\alpha^*_\mathrm{spatula}$ with peeling angle $\theta_\mathrm{p}$ for different strip thicknesses $h$ (``Type I" simulations).}   	 \label{fig:comp_h} 	
	    	\end{center}
	    \end{figure}
	    
	    \begin{figure}[h!]
	    	\hspace{-0.2cm}
	    	\begin{center}
	    		\includegraphics[scale=0.3]{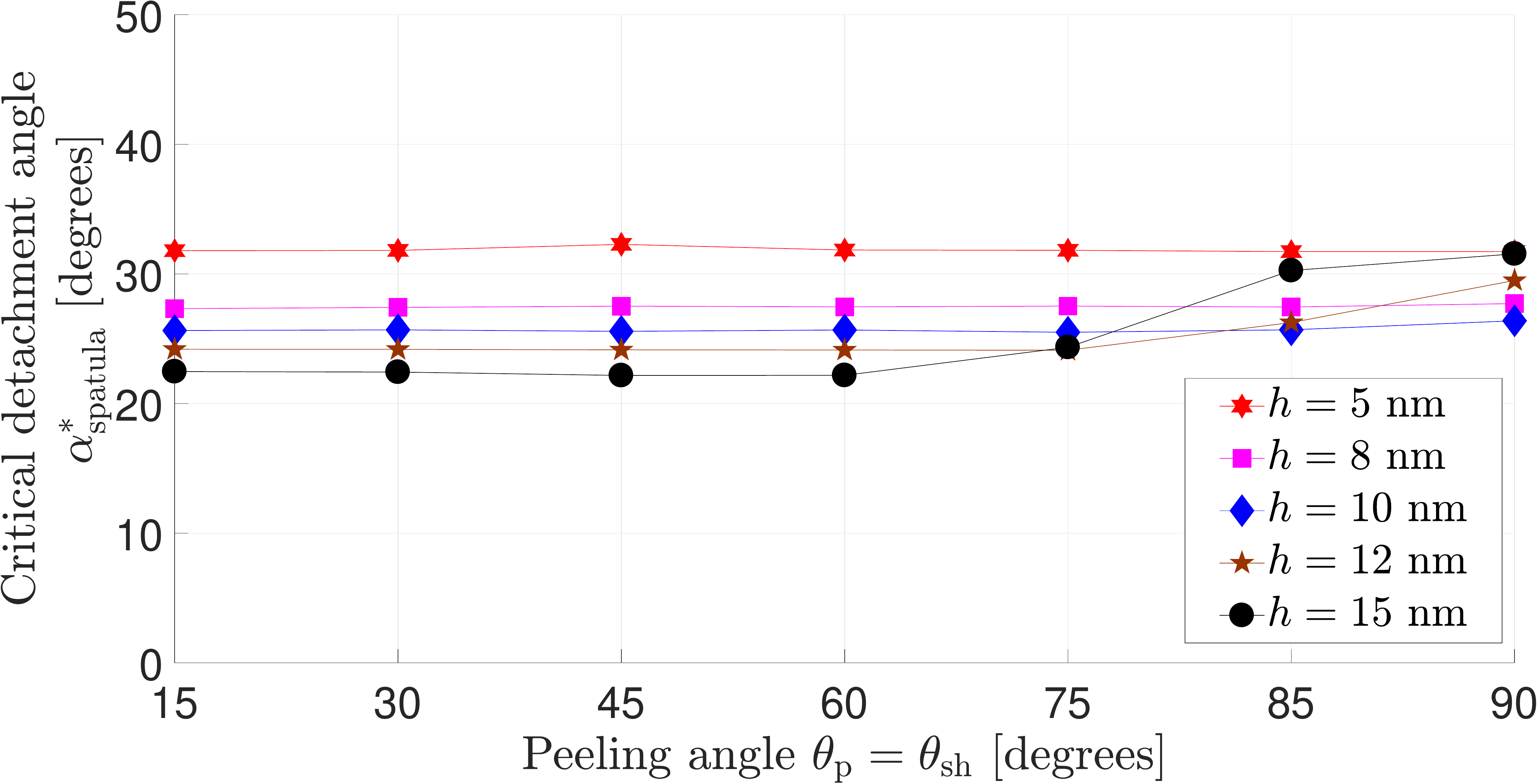} 
	    		\caption{Variation of the critical detachment angle $\alpha^*_\mathrm{spatula}$ with peeling angle $\theta_\mathrm{p} = \theta_\mathrm{sh}$ for different strip thicknesses $h$ (``Type III" simulations).}   	 \label{fig:h_comp_type3} 	
	    	\end{center}
	    \end{figure}
	    
		In Figure~\ref{fig:comp_h} the variation of the critical detachment angle $\alpha^*$ with peeling angle $\theta_\mathrm{p}$ for different spatula pad thicknesses $h$ is shown. It can be observed that, in general, as the thickness decreases the spatula detaches at higher angles. This trend is consistent with the theoretical results of Chen et al.\cite{Chen2009}. Our results also agree well with the experimental observations of Schubert et al. \cite{Schubert2008}, that stiffer fibrillar structures detach at lower critical angles than softer structures. 
		
	    \begin{figure}[]
	    	\begin{subfigure}[ht]{\linewidth}
	    		\begin{center}
	    			\includegraphics[scale=0.31]{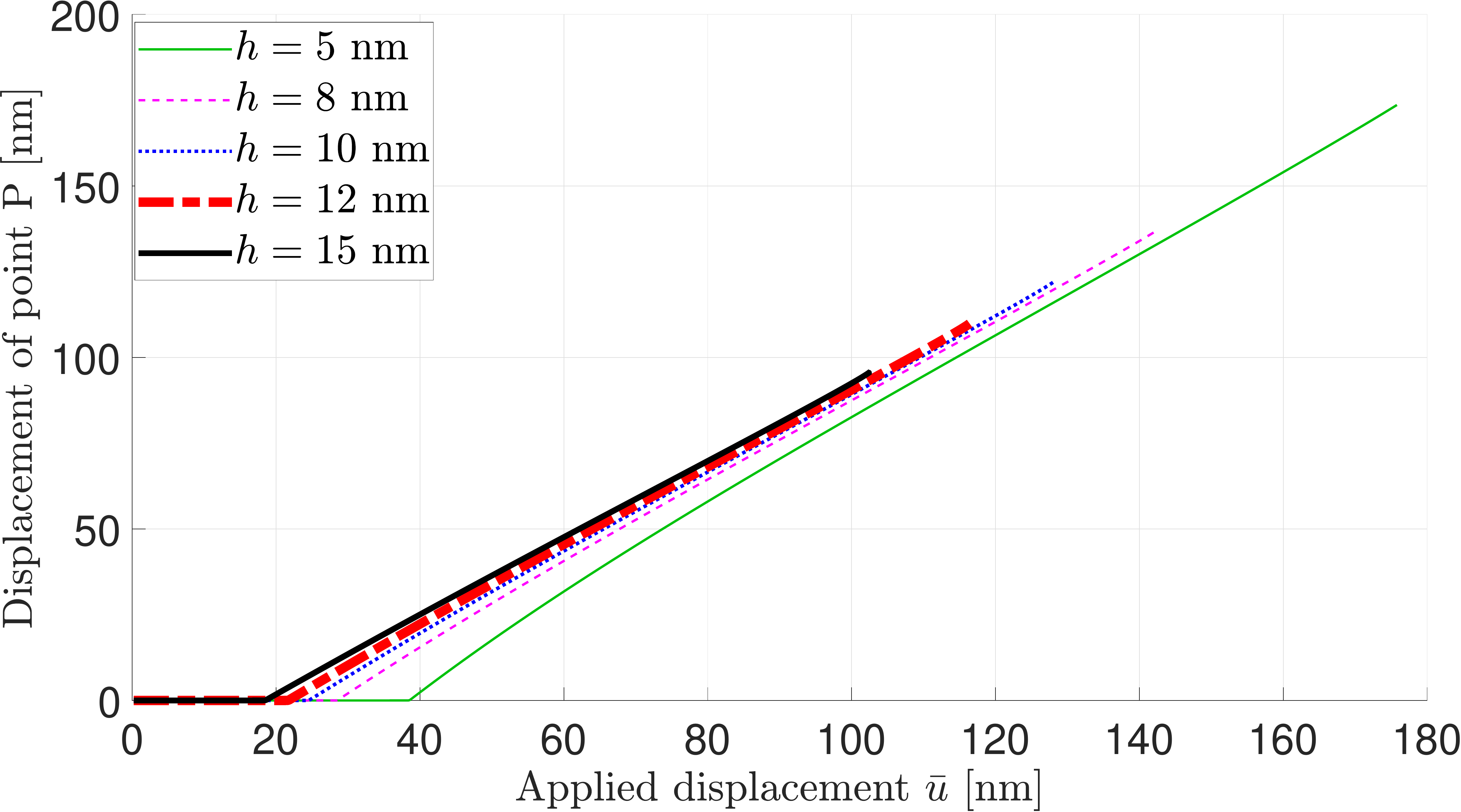}
	    			\caption{$\theta_\mathrm{p} = \theta_\mathrm{sh} = 30^\circ$}
	    			\label{fig:slide_30}
	    		\end{center}
	    	\end{subfigure}\\
	    	\begin{subfigure}[]{\linewidth}
	    		\begin{center}
	    			\includegraphics[scale=0.31]{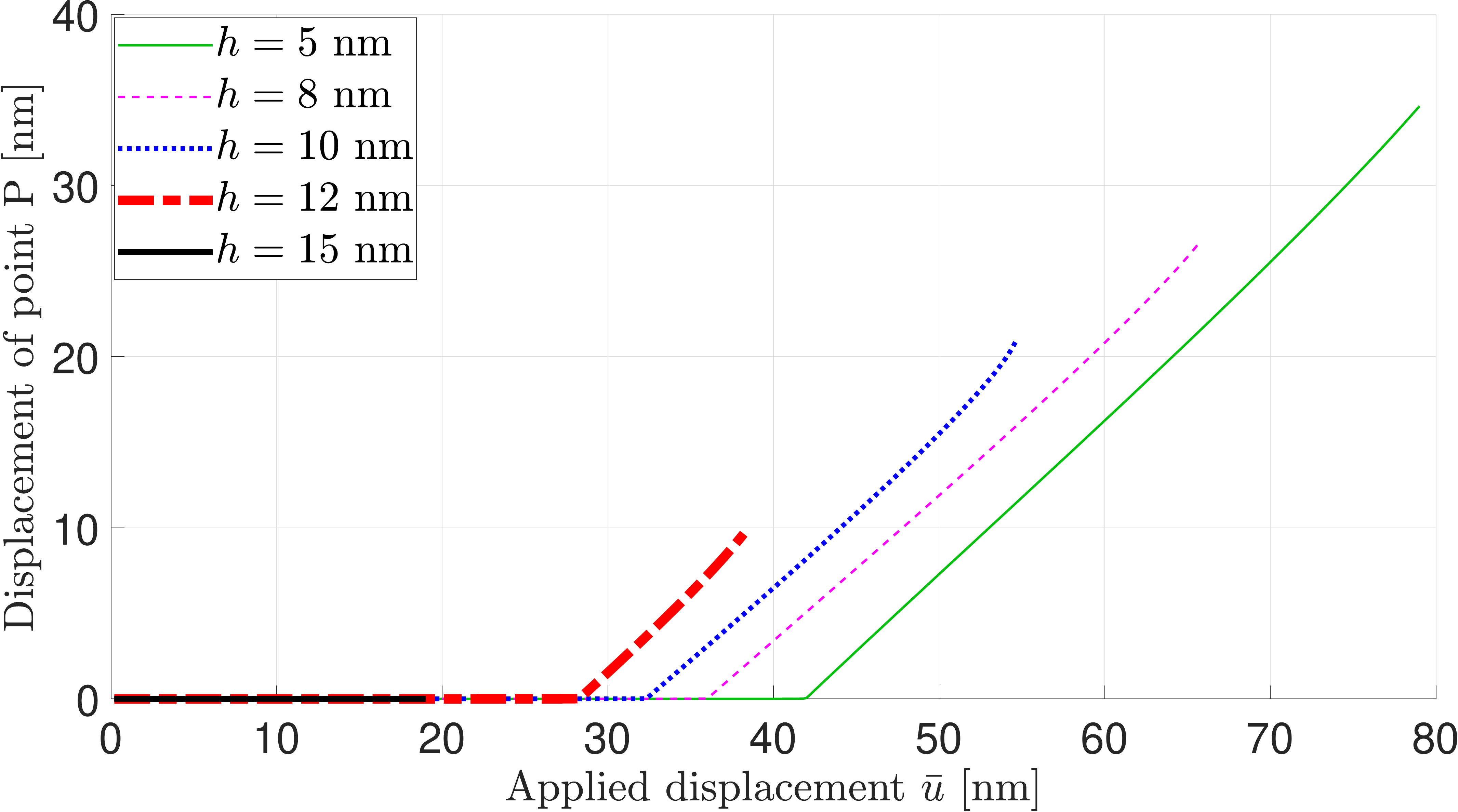}
	    			\caption{$\theta_\mathrm{p} = \theta_\mathrm{sh} = 75^\circ$}
	    			\label{fig:slide_75}
	    		\end{center}
	    	\end{subfigure} \\
	    \end{figure}
    	\begin{figure}\ContinuedFloat
    		\begin{subfigure}[b]{\linewidth}
    			\begin{center}
    				\includegraphics[scale=0.31]{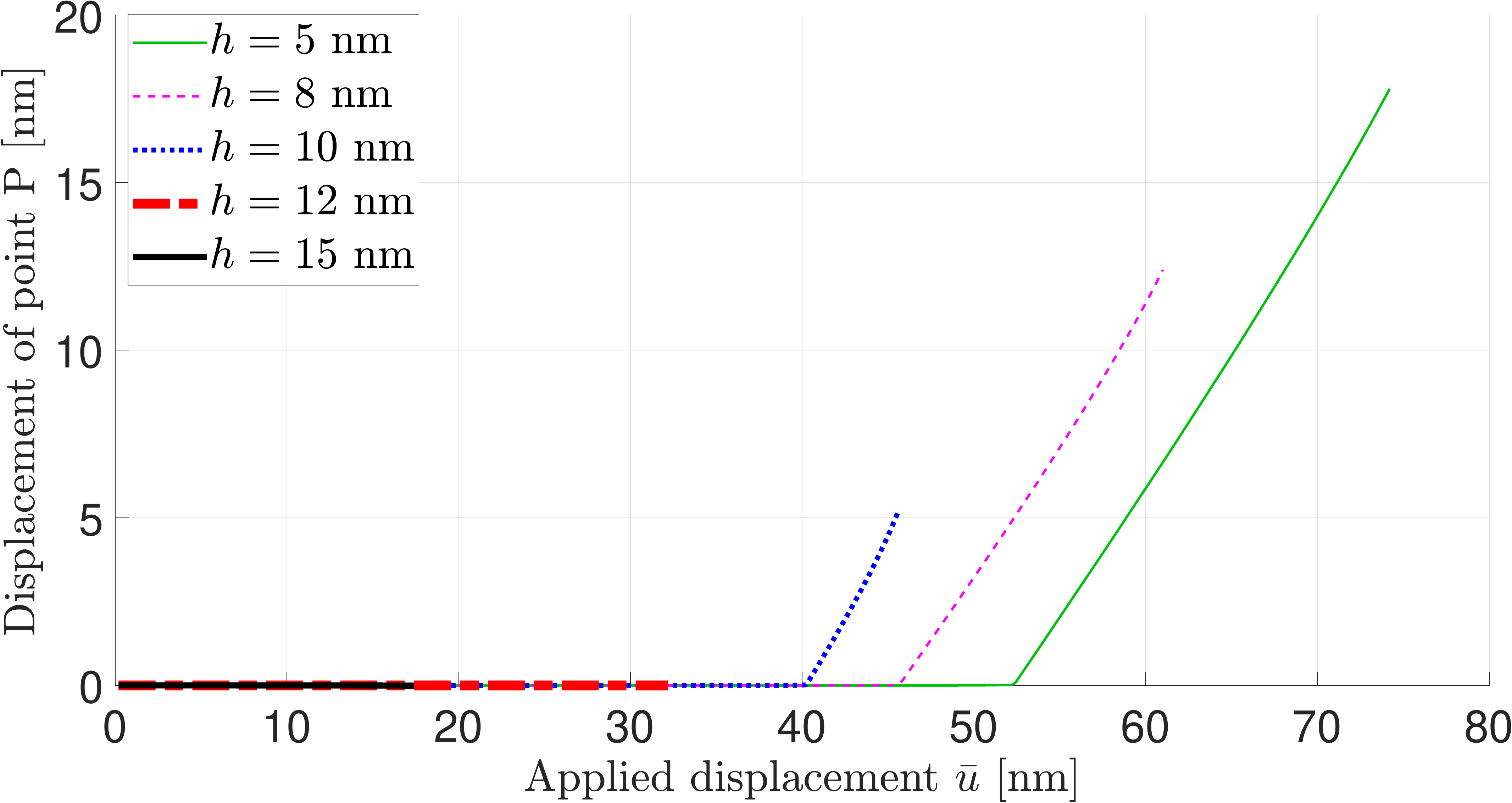}
    				\caption{$\theta_\mathrm{p} = \theta_\mathrm{sh} = 85^\circ$}
    				\label{fig:slide_85}
    			\end{center}
    		\end{subfigure}     
    		\caption{Horizontal displacement of strip point P (see Figure~\ref{fig:strip_orig}) with the applied displacement $\bar{u} = \Vert \bar{\boldsymbol{u}} \Vert$ at different peeling angles $\theta_\mathrm{p}=\theta_\mathrm{sh}$ for different values of strip thickness $h$ (``Type III" simulations).}
    		\label{fig:sliding}
    	\end{figure} 
    
	    However, for ``Type III" peeling simulations (see Figure~\ref{fig:h_comp_type3}) where the spatula is peeled from a pre-rotated configuration (see section~\ref{sec:peel_appl}), the behavior of the spatula to detach at a constant angle ($=\alpha^*$) irrespective of the peeling angle $\theta_\mathrm{p} = \theta_\mathrm{sh}$, is observed only for thickness values of $5-10$\,nm. For large thickness values i.e, $h = 12$ and $15$\,nm, it is found that the spatula does not detach at a constant angle for all the peeling and the shaft angles. Instead, for a large value of the spatula pad thickness, at high angles, i.e., $\theta_\mathrm{p} \geq 75^\circ$, the detachment angle is considerably higher than that for low peeling angles, i.e., $\theta_\mathrm{p} \leq 60^\circ$. 
    
	    In order to understand this behavior, the horizontal displacement of point P (see Figs.~\ref{fig:strip_orig} and ~\ref{fig:strip_rot}) as a function of the applied displacement $\bar{u}$ is plotted in Figure~\ref{fig:sliding} for different peeling angles $\theta_\mathrm{p} = \theta_\mathrm{sh}$ and spatula pad thicknesses. It can be observed that for large spatula pad thicknesses, large peeling and shaft angles, there is still no full sliding of the spatula after the force maximum is reached. For $\theta_\mathrm{p} = 85^\circ$ sliding of the spatula is not observed for both $h = 12$ and $15$\,nm. Correlating these results with those in Figure~\ref{fig:h_comp_type3}, it can be observed that full sliding of the spatula is needed for an invariant critical detachment angle $\alpha^*_\mathrm{spatula}$. 
	    
	    The electron microscopy analyses of Rizzo et al. \cite{Rizzo2006} revealed that the spatula pad of Tokay Gecko is only around $10$\,nm thick. Persson and Gorb \cite{Persson2003} also suggested that the spatula pad thickness is approximately $5-10$\,nm, making it compliant enough to adhere to any kind of substrate. This has also been confirmed by Sauer and Holl \cite{Sauer2013} through detailed 3D finite element simulations. From the results in Figure~\ref{fig:h_comp_type3}, it can be concluded that the thickness of the spatula pad should be small enough to attain constant critical detachment angle. At the same time, as argued by Persson and Gorb \cite{Persson2003} the pad thickness should be large enough to provide sufficient stiffness. In this regard, $h =10$\,nm seems to be an optimum value, i.e., it is the largest value of $h$ with invariant $\alpha^*_\mathrm{spatula}$. 
	    
	    \section{Conclusions}
	    The peeling behavior of a gecko spatula is studied using a coupled adhesion-friction model within the nonlinear finite element method. The spatula, modeled as a thin, two-dimensional strip, is shown to detach at a critical detachment angle that is constant, i.e. it is independent of the peeling angle and the shaft angle. It is also shown that the spatula exhibits the behavior known as ``frictional adhesion" in the literature, according to which the normal adhesion is limited by the frictional force and the critical detachment angle. This critical detachment angle, in general, decreases with increasing spatula pad thickness. However, it is observed that for $h>10$ nm, the spatula does not detach at the same angle for large peeling and shaft angles. For $h>10$ nm, it is observed that, for large peeling and shaft angles, the spatula does not slide on the substrate and this is found to influence the invariance of the critical detachment angle. As such, $h=10$ nm can be regarded as the optimum value with respect to invariant critical detachment angle.  
	    
	    \section*{Acknowledgement}
	    The authors are grateful to the SERB, DST for supporting this research under project SR/FTP/ETA-0008/2014. The authors thank Dr. David Labonte for his valuable comments.

	    \end{document}